\documentclass[twocolumn,english,aps,prb,showpacs,showkeys,notitlepage,superscriptaddress,noeprint]{revtex4-2}
\ifdefined\pdfoutput
	\pdfoutput=1
\fi
\usepackage{tikzexternal}

\RequirePackage{hyperref}
\hypersetup{colorlinks=true,linkcolor=blue,citecolor=red}

\usepackage{natbib}

\usepackage{xcolor}
\definecolor{MATLAB1}{HTML}{0072BD}
\definecolor{MATLAB2}{HTML}{D95319}
\definecolor{MATLAB3}{HTML}{EDB120}
\definecolor{MATLAB4}{HTML}{7E2F8E}
\definecolor{MATLAB5}{HTML}{77AC30}
\definecolor{MATLAB6}{HTML}{4DBEEE}
\definecolor{MATLAB7}{HTML}{A2142F}

\usepackage{amsmath}
\usepackage{amssymb}
\usepackage{mathtools}

\ifdefined\siunitxMode
	\usepackage{siunitx}
\else
	\newcommand{\SI}[2]{#1\,#2}
	
	\newcommand{\num}[1]{#1}
	\newcommand{\us}{\mu\text{s}}
	\newcommand{\MHz}{\text{MHz}}
	
	\newcommand{\percent}{\%}
	\newcommand{\radian}{\text{rad}}
	\newcommand{\micro}{\mu}
	\newcommand{\electronvolt}{\text{eV}}
	
\fi

\usepackage{physics}
\usepackage{microtype}

\usepackage[capitalise]{cleveref}
\Crefname{section}{Sec.}{Secs.}

\usepackage{booktabs}
\usepackage{iftex}
\usepackage{leftindex}
\RequirePackage[normalem]{ulem} % strikethrough

\ifxetex
	\usepackage{unicode-math}
	\usepackage{fontspec}
	\selectfont
\else
	\usepackage{lmodern}
\fi

\makeatletter
\AtBeginDocument{\let\LS@rot\@undefined}
\makeatother

\newcommand{\ct}[1]{#1^\dagger}
\newcommand{\conj}[1]{#1^\ast}

\newcommand{\ii}{i}
\newcommand{\ee}{e}

\newcommand{\sumjInAB}[1]{\sum_{\mathclap{j\in\qty{A,B}}}{#1}}
\newcommand{\aHat}{\hat{a}}
\newcommand{\actHat}{\ct{\aHat}}
\newcommand{\tauXinit}{\tilde{\tau}_x}

\newcommand{\tauZinit}{\tilde{\tau}_z}

\newcommand{\tauXcharge}{\tau_x}

\newcommand{\tauZcharge}{\tau_z}

\newcommand{\tauZqubit}{\hat{\tau}_z}
\newcommand{\tauMqubit}{\hat{\tau}_-}
\newcommand{\tauPqubit}{\hat{\tau}_+}
\newcommand{\sigmaXinit}{\sigma_x}

\newcommand{\sigmaZinit}{\sigma_z}
\newcommand{\sigmaMinit}{\sigma_-}
\newcommand{\sigmaPinit}{\sigma_+}

\newcommand{\sigmaZqubit}{\hat{\sigma}_z}
\newcommand{\sigmaMqubit}{\hat{\sigma}_-}
\newcommand{\sigmaPqubit}{\hat{\sigma}_+}
\newcommand{\Heff}{\hat{H}_\text{eff}}
\newcommand{\lindOp}[2]{\mathcal{D}\bqty{#1}#2}
\newcommand{\measOp}[2]{\mathcal{M}\bqty{#1}#2}
\newcommand{\varLabel}[2]{#1^{\pqty{#2}}}

\newcommand{\wx}{b_x}

\newcommand{\wz}{B_z}
\newcommand{\wc}{\omega_c}
\newcommand{\wdr}{\omega_d}
\newcommand{\wsp}{\omega_\mathrm{sp}}
\newcommand{\wch}{\omega_\mathrm{ch}}
\newcommand{\gc}{g_c}
\newcommand{\tc}{t_c}

\newcommand{\Dc}{\Delta_c}
\newcommand{\Dch}{\Delta_\mathrm{ch}}
\newcommand{\Dsp}{\Delta_\mathrm{sp}}

\newcommand{\gammaOne}{\gamma_1}
\newcommand{\gammaPhi}{\gamma_\phi}

\newcommand{\dt}{\dd{t}}

\newcommand{\dWt}{\dd{W\pqty{t}}}

\newcommand{\Ohat}{\hat{\mathcal{O}}}
\newcommand{\VP}{V_P}
\newcommand{\VPt}{V_P\pqty{t}}
\newcommand{\Vint}{V_\text{int}}

\newcommand*{\ketSub}[2]{{\ket{#1}}_{\!#2}}
\newcommand*{\braSub}[2]{\leftindex_{#2\!}{\bra{#1}}}

\newcommand*{\driveAmp}{p}

\newcommand*{\avgLowKappaFid}{61}
\newcommand*{\maxHighKappaFid}{98}
\newcommand*{\avgHighKappaFid}{81}

\begin{document}

\title{Measurement-Based Entanglement of Semiconductor Spin Qubits}

\author{Remy L. Delva}
\affiliation{Department of Physics, Princeton University, Princeton, New Jersey 08544, USA}
\author{Jonas Mielke}
\affiliation{Department of Physics, University of Konstanz, D-78457 Konstanz, Germany}
\author{Guido Burkard}
\affiliation{Department of Physics, University of Konstanz, D-78457 Konstanz, Germany}
\author{Jason R. Petta}
\affiliation{Department of Physics and Astronomy, University of California, Los Angeles, Los Angeles, California 90095, USA}
\affiliation{Center for Quantum Science and Engineering, University of California, Los Angeles, Los Angeles, California 90095, USA}

\begin{abstract}
Measurement-based entanglement is a method for entangling quantum systems through the state projection that accompanies a parity measurement.
We derive a stochastic master equation describing measurement-based entanglement of a pair of silicon double-dot flopping-mode spin qubits, develop numerical simulations to model this process, and explore what modifications could enable an experimental implementation of such a protocol.
With device parameters corresponding to current qubit and cavity designs, we predict an entanglement fidelity \(F_e\approx\avgLowKappaFid\%\). By increasing the cavity outcoupling rate by a factor of ten, we are able to obtain a simulated \(F_e\approx\avgHighKappaFid\%\)  while maintaining a yield of 33\%.
\end{abstract}

\maketitle

\section{Introduction}

Due to their long spin-coherence times and reliance on proven nanofabrication technologies, electron spin qubits in silicon are strong candidates for use in a quantum computer \cite{Burkard2023}.
Spins generally interact through the exchange interaction, which is based on electrical control of wave function overlap \cite{Petta2005}.
Previous implementations of two-qubit gates in silicon quantum devices have used tunable exchange couplings to establish interactions between neighboring qubits; these interactions can evolve an unentangled two-qubit state into an entangled state \cite{Zajac2018,Watson2018,Noiri2022,Xue2022,Mills2022}.
However, for nonadjacent qubit pairs, this type of gate is limited by the short effective range of the exchange interaction.

Circuit quantum electrodynamics (cQED) is a device architecture that has enabled strong coupling between microwave frequency photons and superconducting qubits \citep{Blais2004,Wallraff2004}.
Long-distance coupling of superconducting qubits has also been achieved with cQED \cite{Majer2007,Sillanpaeae2007}.
Efforts have been made to broaden cQED by incorporating semiconductor quantum dots in microwave cavities \cite{Cottet2010,Frey2012,Petersson2012}; strong spin-photon coupling, resonant spin-spin interactions, and dispersive spin-spin coupling have all been experimentally demonstrated \cite{Mi2018,Samkharadze2018,Landig2018,Borjans2019,HarveyCollard2022,dijkema2023}.
Further improvements in device performance may enable the generation of highly entangled spin states with quantum dot cQED.

We can consider two classes of techniques used to generate entanglement in cQED device architectures.
The first is to implement a non-local entangling gate, such as an iSWAP, by precisely controlling the length of the interaction.
An iSWAP gate transforms \(\ket{\downarrow\uparrow}\mapsto -\ii\ket{\uparrow\downarrow}\) and \(\ket{\uparrow\downarrow}\mapsto -\ii\ket{\downarrow\uparrow}\), leaving \(\ket{\downarrow\downarrow}\) and \(\ket{\uparrow\uparrow}\) unchanged \citep{Krantz2019,Young2022}.
An alternate means of generating entanglement is to exploit the properties of quantum measurement to project an initially unentangled two-qubit state onto an entangled subspace \citep{Cabrillo1999}.
The second approach is generally refered to as measurement-based entanglement (MBE).

As an example of an MBE protocol, suppose we prepare two qubits in the product state:
\begin{equation}
	\ket{\Psi_{\rm init}} = \ket{+x}\ket{+x}
	=\frac{1}{2}\qty\big(\ket{\downarrow\downarrow}+\ket{\downarrow\uparrow}+\ket{\uparrow\downarrow}+\ket{\uparrow\uparrow}).
	\label{eqn:prodState}
\end{equation}

A non-demolition parity measurement of the two-qubit system will produce one of the following:
\begin{alignat}{5}
	\text{Even Parity}&\Rightarrow&&\ket{\Phi_+}&&=\frac{1}{\sqrt{2}}\bigl(\ket{\downarrow\downarrow}&&+\ket{\uparrow\uparrow}&&\bigr),\\
	\text{Odd Parity}&\Rightarrow&&\ket{\Psi_+}&&=\frac{1}{\sqrt{2}}\bigl(\ket{\downarrow\uparrow}&&+\ket{\uparrow\downarrow}&&\bigr).
\end{alignat}
By repeating this process of initialization and measurement while postselecting on the measured parity, we can obtain a specific two-qubit Bell state.
Alternately, we can condition the application of a single-qubit Pauli $x$-gate upon the measurement of the undesired parity to obtain a specific Bell state deterministically.
MBE protocols of the former kind have previously been demonstrated with trapped ions \citep{Lanyon2013}, transmon qubits \citep{Riste2012}, and nitrogen vacancy centers in diamond \citep{Pfaff2012}.
Additionally, experiments with transmon qubits have implemented protocols of the latter kind, in which states of unwanted parity are rotated onto the desired state using unitary gates \citep{Riste2013}.

Compared to implementations utilizing unitary qubit-qubit interactions, the MBE approach has the potential to be less demanding in terms of dynamical fine-tuning, since during the measurement process the state will nominally approach one of the states corresponding to a measurement outcome, with a minimum of oscillatory behavior.
The lack of direct coupling may also help to prevent undesirable qubit-qubit interactions during other stages of a quantum algorithm: effective interactions can be eliminated by simply turning off the measurement signal, without requiring any retuning of qubit parameters.
Extensions of the parity-measurement protocol to more than two qubits can also enable the generation of Greenberger–Horne–Zeilinger (GHZ) states with fewer operations, albeit non-deterministically; if conditional unitary corrections are also implemented, MBE can produce such states deterministically \citep{Riste2013}.

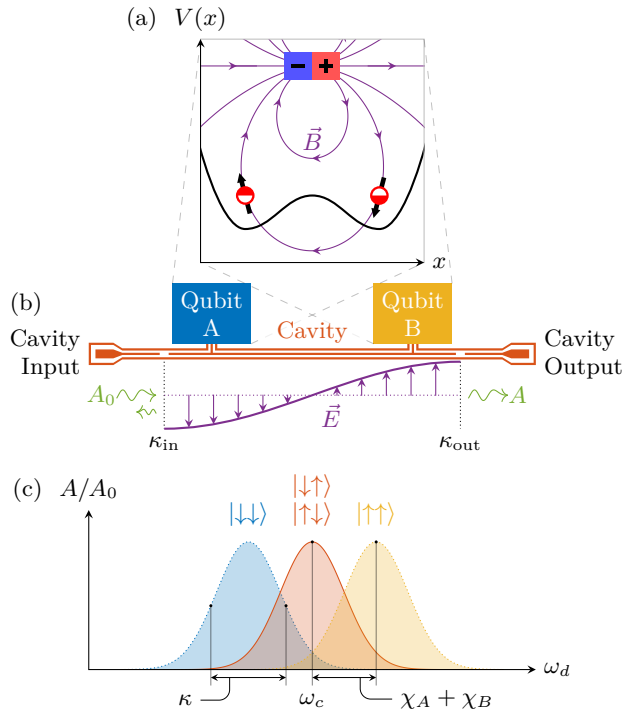
\begin{figure}
	{
		\newcommand*{\figWidth}{\textwidth}
		\tikzsetnextfilename{cQEDdiagram}
		\begin{tikzpicture}[x=\columnwidth*11/16,y=\columnwidth*11/16]
			
		\end{tikzpicture}
	}
	\protect\caption
	{
		\label{fig:mbe}(a) Illustration of the spin-charge coupling mechanism within a flopping-mode DQD.
		The presence of a gradient magnetic field \(\vec B\) causes the Zeeman splitting and quantization axis of the electron spin to depend on its position.
		This allows the electric field \(\vec E\) of the microwave cavity to couple to the spin of the electron.
		(b) Schematic of a MBE device.	Qubits \(A\) and \(B\) are located at the anti-nodes of a half-wavelength (\(\lambda/2\)) cavity with input and output couplings \(\kappa_\text{in}\) and \(\kappa_\text{out}\).
		(c) Illustration of the normalized cavity transmission \(A/A_0\) as a function of cavity drive frequency \(\wdr\) for a generic MBE protocol using two-level qubits.
		The bare cavity-resonance frequency \(\wc\) is shifted by either \(\pm\qty(\chi_A + \chi_B)\) for even-parity states, or zero for odd-parity states (assuming \(\chi_A = \chi_B\)).
		Probing the cavity transmission at \(\omega_c\) therefore implements a parity measurement.
		\message{Caption font size is: \the\font}
	}
\end{figure}

In this article we evaluate the feasibility of using dispersive or near-dispersive cQED measurements to implement MBE protocols with semiconductor spin qubits.
Previous analyses of superconducting
\citep{Cabrillo1999,Hutchison2009,Lalumiere2010,Pfaff2012,Lanyon2013,Riste2013}
and semiconductor \cite{Zhu2023} qubit MBE protocols have treated the qubits as ideal two-level systems.
For cavity-coupled semiconductor double quantum dot (DQD) flopping-mode spin qubits, spin-photon coupling results from a combination of electric dipole coupling of a single electron charge to the cavity electric field and synthetic spin-orbit coupling produced by a magnetic field gradient, as shown in \cref{fig:mbe}(a). As such, the DQD is described by a four-level system consisting of hybridized orbital and spin states \cite{Benito2017}.
We model the evolution of a system consisting of two cavity-coupled flopping-mode spin qubits that are subjected to a continuous homodyne parity measurement.
Section \ref{ch:cQED_MBE_intro} characterizes the performance of MBE protocols for two-level qubits that are subjected to parity measurements.
In \cref{ch:DQD_MBE} we describe the Hamiltonian governing a system consisting of two cavity-coupled flopping-mode spin qubits. Section \ref{ch:SimResults} characterizes the fidelity of spin-qubit MBE protocols, comparing entanglement fidelities that can be obtained with existing device parameters. We also estimate the device parameters that will be needed to achieve postselected fidelities \(F_{\ket{\Psi_+}} > \SI{80}{\percent}\) to the Bell state \(\ket{\Psi_+}\). We conclude in \cref{ch:Conclusion} with a summary of practical alterations to existing cavity designs and an outlook for further work on this subject.

\section{\label{ch:cQED_MBE_intro}MBE for two-level systems}

We first review how a MBE protocol might be implemented for a cQED system using an idealized Hamiltonian. For two-level qubits \(A\) and \(B\) coupled to a resonant cavity, as shown in \cref{fig:mbe}(b), the standard Jaynes-Cummings Hamiltonian is
\begin{equation}
	\begin{split}
		H ={}& \hbar \wc\actHat\aHat + \hbar \sum_{\mathclap{j\in\qty{A,B}}\hspace{1em}}\qty[\frac{\omega_j}2\varLabel{\sigmaZinit}j + g_j\qty(\actHat\varLabel{\sigmaMinit}j+\aHat\varLabel{\sigmaPinit}j)],
	\end{split}
\end{equation}
where \(\wc\) is the resonance frequency of the cavity, \(\omega_{A\qty(B)}\) are the qubit transition frequencies, and \(g_{A\qty(B)}\) are the qubit-cavity coupling rates. Unless specified otherwise, all frequencies are given in units of rad/s.
\(\varLabel{\sigma_{z,+,-}}{A\qty(B)}\) are the standard Pauli operators applied to the relevant qubit, and \(\aHat\)(\(\actHat\))  is the cavity photon annihilation(creation) operator.

In the dispersive regime, i.e.\ where \(g_{A\qty(B)} \ll \Delta_{A\qty(B)}\), the Hamiltonian can be approximated as \citep{Blais2007},
\begin{equation}
	\label{eqn:Hdisp}
	\begin{split}
		H_\text{disp} ={}& \hbar \qty(\wc+\sumjInAB{\chi_j\varLabel{\sigma_z}{j}})\actHat\aHat +  \sum_{\mathclap{j\in\qty{A,B}}\hspace{1em}}{\frac{\hbar}{2}\qty(\omega_j + \chi_j)\varLabel{\sigma_z}{j}}\\
		{}&+ \hbar J\qty(\varLabel{\sigma_{\smash{-}}}{A}\varLabel{\sigma_{\smash{+}}}{B} + \varLabel{\sigma_{\smash{+}}}{A}\varLabel{\sigma_{\smash{-}}}{B}),
	\end{split}
\end{equation}
where \(J=\tfrac{g_Ag_B\qty(\Delta_A + \Delta_B)}{2\Delta_A\Delta_B}\), \(\Delta_j = \omega_j - \omega_c\) are qubit-cavity detunings, and \(\chi_j = \frac{g_j^2}{\Delta_j}\) are the dispersive shifts for the two qubits \citep{Hutchison2009}.

The first term in \cref{eqn:Hdisp} implies that the cavity resonance frequency is dispersively shifted by an amount \(\sum_{j\in\qty{A,B}}{\chi_j\varLabel{\sigmaZinit}{j}}\) that is dependent on the state of the qubits.
Now suppose that the qubit-cavity couplings are set such that \(g_A=-g_B=g\), and the qubit-cavity detunings satisfy \(\Delta_A = \Delta_B = \Delta\).
In this case, the dispersive shifts induced by both qubits are equal.
Then, if both qubits are in the ground state such that  \(\ketSub{\psi}{AB}=\ketSub{\downarrow}{A}\ketSub{\downarrow}{B}\), the cavity resonance will be dispersively shifted down in frequency. Similarly, if \(\ketSub{\psi}{AB} = \ketSub{\uparrow}{A} \ketSub{\uparrow}{B}\), the cavity resonance will be shifted up in frequency.  These two cases are shown as the blue and yellow curves in \cref{fig:mbe}(c), respectively.
In contrast, if the qubits are in opposite states the two dispersive shifts will cancel out, and the cavity response will be unchanged, as shown by the orange curve in \cref{fig:mbe}(c). Measurements of the normalized cavity transmission $A/A_0$ performed with the cavity drive frequency \(\omega_d\) tuned to the cavity frequency (i.e. with the cavity-drive detuning  $\Delta_c = \omega_c - \omega_d$ = 0), will depend on the parity of the two qubits.

Two issues must be noted before modeling the parity-measurement process. Firstly, any real qubit will be subject to a noisy environment, causing dephasing and other incoherent evolution to occur at some rate.
The cavity measurement process is not instantaneous, and the assorted qubit and cavity detunings and coupling strengths will similarly establish a rate at which measurement (and the corresponding projection) occurs.
A successful experimental implementation must use a set of realistic device and measurement parameters that will allow the qubits to be entangled more quickly than they will decohere; therefore our simulation must model both measurement and decoherence as continuous phenomena.

Secondly, depending on the regime in which the qubits are operated, the cavity may have other effects on the qubits. The third term in \cref{eqn:Hdisp}, corresponding to coherent qubit-qubit interactions mediated by virtual cavity photons, is an example of such an effect. Resonant exchange of excitations between a qubit and the cavity also becomes possible if the device leaves the dispersive regime, i.e. if \(g \gtrsim \Delta\).

Optimizing the qubit and measurement parameters is therefore crucial for the success of the experiment; the capacity to accurately simulate the influences of various measurement parameters on the final distribution of states is consequently an important step towards an effective experimental realization of MBE with silicon spin qubits.

For a system of two ideal two-level qubits dipole-coupled to a driven cavity, we obtain the following  master equation describing the evolution of the combined system in time, with respect to a reference frame obtained by first transforming to the rotating reference frame defined by 
\begin{equation}
	U = \exp[\ii\qty(\sum_{\mathclap{\hspace{1.75em}j\in\{A,B\}}}{\frac{\omega_j\varLabel{\sigmaZinit}{j}}2 + \wdr\actHat\aHat})t],
	\label{eqn:Utrans}
\end{equation}
and subsequently transforming to the displaced frame set by \(D\qty[\alpha]=\exp(\alpha\actHat-\conj\alpha\aHat)\):
\begin{equation}
	\label{eqn:mastEqTwoLevel}
	\begin{split}
		\dd{\rho} = \mathcal{L}\rho\dt ={}& -\frac{\ii}{\hbar}\comm{\Heff}{\rho}\dt\\
		&+\sum_{\mathclap{j\in\Bqty{A,B}\hspace{1em}}}\pqty{\varLabel{\gammaOne}{j}\lindOp{\varLabel{\sigmaMinit}{j}}{\rho}+\frac{\varLabel{\gammaPhi}{j}}{2}\lindOp{\varLabel{\sigmaZinit}{j}}{\rho}}\dt\\
		&+\kappa\qty\big(\lindOp{\aHat}{\rho})\dt.
	\end{split}
\end{equation}

Here \(\mathcal L\) is a superoperator for the deterministic evolution of \(\rho\) and $\mathcal{D}[\mathcal{O}]$ is the dissipation superoperator
 
 \begin{equation}
\mathcal{D}[\mathcal{O}] \rho = \mathcal{O} \rho \ct{\mathcal{O}} - \frac12 \acomm{\ct{\mathcal{O}}\mathcal{O}}{\rho}.
 \end{equation}

\noindent The effective Hamiltonain is obtained by consecutively transforming $H$ [Eq. (4)] to the frame set by $U$ [Eq. (6)] and $D[\alpha]$,
\begin{equation}
	\label{eqn:twoLevelHeff}
	\begin{split}
		\hat H_\text{eff} ={}& \hbar \Dc\actHat\aHat \\
		&+ \hbar \sum_{\mathclap{j\in\qty{A,B}\hspace{1em}}}\Bigl[
		\begin{aligned}[t]
			&g_j\ee^{\ii\Delta_jt}\qty(\aHat-\alpha)\varLabel{\sigmaPinit}{j}\\
			&+g_j\ee^{-\ii\Delta_jt}\qty(\actHat-\conj\alpha)\varLabel{\sigmaMinit}{j}\smash[t]{\Bigr]},
		\end{aligned}
	\end{split}
\end{equation}

with 
\begin{equation}
	\label{eqn:twoLevelAlphaDot}
	\dot\alpha = -\ii\Delta_c\alpha + \ii\sqrt{\kappa_\text{in}}\driveAmp\qty(t) -\frac{\kappa}{2}\alpha.
\end{equation}
Here \(\alpha\) is the cavity coherent state population, \(\kappa\) = \(\kappa_\text{in} + \kappa_\text{out}\), \(\driveAmp\qty(t)\) is the amplitude of the cavity drive in units of \(\text{photons}/\text{time}\), and \(\varLabel{\gammaOne}j\)(\(\varLabel{\gammaPhi}j\)) is the relaxation(dephasing) rate of qubit $j$. The terms in \cref{eqn:mastEqTwoLevel} account for the system's coherent evolution, qubit relaxation and dephasing, and cavity loss, respectively. Note that this master equation does not make a dispersive assumption, resulting in the effective Hamiltonian being distinct from \cref{eqn:Hdisp}.

While \cref{eqn:mastEqTwoLevel} models the system's incoherent behavior, it only provides the unconditioned evolution of the system: the dependence of the final state on the observed measurement outcome is not captured.
In order to associate measurement outcomes with final values of the system's density matrix, we must add a fourth, stochastic term.
We can then numerically simulate many specific evolutions of the density matrix, recording the final state of the system and experimental measurement result for each iteration.

A continuous homodyne measurement of the microwave transmission through the cavity described by a measurement efficiency \(\eta\) and phase offset \(\phi\) adds a stochastic term to the master equation \citep{Wiseman2009}.
Denoting the conditional density matrix that evolves according to this stochastic differential equation \(\rho_{\text{cond.}}\), we obtain
\begin{equation}
	\label{eqn:masterEqStoch}
	\dd{\rho_{\text{cond.}}} = \mathcal{L}\rho_{\text{cond.}}\dt + \sqrt{\kappa_\text{out}\eta}\measOp{\ee^{\ii\phi}\aHat}{\rho_\text{cond.}}\dWt,
\end{equation} with
\begin{equation}
\measOp{\Ohat}{\rho} = \Ohat\rho + \rho\ct{\Ohat} - \expval{\Ohat + \ct{\Ohat}}\rho.
\label{eqn:VpStoch}
\end{equation}
Here \(\dWt\) is a stochastic variable with zero mean and variance \(\dt\).
For each sampling (or ``unraveling") of \(\dWt\), we can then calculate the resulting homodyne output signal \(\VP\),
\begin{equation}
	\VPt\dt \propto \sqrt{\kappa_{\rm out}\eta}\expval{\ee^{-\ii\phi}\actHat + \ee^{\ii\phi}\aHat} \dt + \dWt.
\end{equation}
Since the physical homodyne output will depend on experimental
specifics such as the amplifier gain, we will use arbitrary units for
  \(\VPt\)   in the following sections.

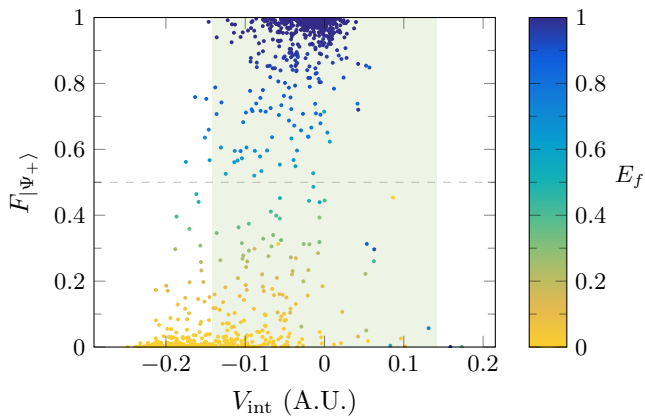
\begin{figure}
	\tikzsetnextfilename{HutchScatter}
	\begin{tikzpicture}
		
	\end{tikzpicture}
	\protect\caption
	{
		\label{fig:hutchFidelityVsSignal}
		Scatter plot of the Bell state fidelity \(F_{\ket{\Psi_+}}\) as a function of \(\Vint\) using two-level device parameters from \citealp[Ref.\!~][Figure 4 inset]{Hutchison2009}.
		Defining the measurement rate \(\Gamma_{\mathrm{ci}}/2\pi=\frac{g^2\kappa}{\pi \Delta^2}=\SI{0.032}{\MHz}\), we have \(\wdr/2\pi = \wc/2\pi = \SI{0.796}{\MHz}\), \(\kappa/2\pi = \qty(\frac\Delta g)^2\frac{\Gamma_\mathrm{ci}} {4\pi} = \SI{1.592}{\MHz}\), \(\Delta/2\pi = 5g/\pi = \SI{15.915}{\MHz}\), \(\driveAmp/2\pi = \frac{\Gamma_\mathrm{ci}}{16\sqrt{2}\pi} = \SI{2.814}{\MHz}\), and \(\eta = 1\) (taking \(\kappa_\text{in}/\kappa_\text{out} = 1/8\)).
		This simulation contains 1200 trajectories, shown here at time \(t=5\Gamma_\mathrm{ci}^{-1}=\SI{25}{\us}\).
		The point colors correspond to the final-state entanglement of formation \(E_f\) of each trajectory.
		The shaded green background region indicates the interval where \(\abs{\Vint} \le t\sqrt{\Gamma_{\mathrm{ci}}}\).
		This inequality is used to define which measurements are accepted and included in subsequent calculations of the average entangled-state fidelity (e.g.\ in \cref{fig:hutchConcVsT}).
		The dashed gray line is a guide to the eye, indicating a fidelity of \(\SI{50}{\percent}\).
	}
\end{figure}

The MBE protocol begins with the qubits initialized in the state
$\ket{\Psi_{\rm init}}$ defined in
  Eq.~\eqref{eqn:prodState}.
A constant cavity drive with amplitude \(\driveAmp = \frac{g^2\kappa}{2\sqrt2\Delta^2}\) is applied for the duration of the measurement.
During each measurement, \(\VPt\) is integrated over time to produce an associated scalar value \(\Vint\), which is then used to determine whether the resulting two-qubit state will be retained or discarded.
Within the simulation, each of these measurement sequences corresponds to an unraveling of the stochastic master equation, \cref{eqn:masterEqStoch}, with an independent random sampling of the stochastic variable d$W(t)$. 
The full simulation run evaluates 1200 such unravelings using a stochastic Runge-Kutta method similar to that described in \citealp[Ref.][]{Kloeden1992}.
The entanglement of formation \(E_f\qty[\rho_\text{red.}]\) \citep{Wootters1998} and fidelity to the target Bell state \(F_{\ket{\Psi_+}}=\Tr[\rho_\text{red.}\ketbra{\Psi_+}{\Psi_+}]\) are then calculated for the final density matrix of each unraveling, where \(\rho_\text{red.}\) is the reduced density matrix obtained by taking the partial trace of \(\rho_\text{cond.}\) with respect to the cavity degree of freedom, i.e. \(\rho_{\mathrm{red.}}=\Tr_{\mathrm{cav.}}\left[\rho_{\mathrm{cond.}}\right]\).
These values can then be used to generate scatter plots showing the extent of the correlation between the measurement results (\(\Vint\)) and the final fidelity to the target state (\(F_{\ket{\Psi_+}}\)).

We first simulate the operation of an idealized two-level device evolving according to \cref{eqn:masterEqStoch}. Figure \ref{fig:hutchFidelityVsSignal} shows a scatter plot of $F_{\ket{\Psi_+}}$ as a function of the integrated homodyne output $V_{\rm int}$ for a given trajectory.  Each point corresponds to a single unraveling of \cref{eqn:masterEqStoch}.  The parameters used in the simulation are listed in the \cref{fig:hutchFidelityVsSignal} caption and correspond to those used in previous work describing MBE protocols for superconducting qubits, where \(g = g_A = -g_B\) and \(\Delta = \Delta_A = \Delta_B\) \citep{Hutchison2009}.

\begin{figure}[t!]
	\tikzsetnextfilename{HutchConc}
	\begin{tikzpicture}
		
	\end{tikzpicture}
	\protect\caption
	{
		\label{fig:hutchConcVsT}
		Average concurrence $C$ as a function of time for the same simulation run shown in \cref{fig:hutchFidelityVsSignal}. Only data points from the shaded green region of \cref{fig:hutchFidelityVsSignal} contribute to the average: this effectively limits the average to trajectories in the high-fidelity cluster near \(\Vint = 0\). The dashed green line shows results from Ref.\ \cite{Hutchison2009}, Fig.~4, specifically in the case where \(g^2\kappa/\Delta^2=\Gamma_\mathrm{ci}/2\). The solid orange line shows the results of our non-dispersive simulation of the same two-level system for the same parameters. Note that the oscillatory behavior seen in this plot was not observed in Ref.\ \cite{Hutchison2009}, which used a dispersive approximation in which cavity-mediated qubit-qubit effects were neglected by setting \(J=0\). Inset: An enlarged view of the results for \(t\in\qty[0,2\Gamma_{\mathrm{ci}}^{-1}]\). The light blue line shows the theoretically predicted concurrence according to a dispersive model with fully coherent evolution [see \cref{eqn:concOsc}] and \(J=\frac{g_Ag_B\qty(\Delta_A+\Delta_B)}{2\Delta_A\Delta_B}=-\frac{g^2}{\Delta}\).}
\end{figure}
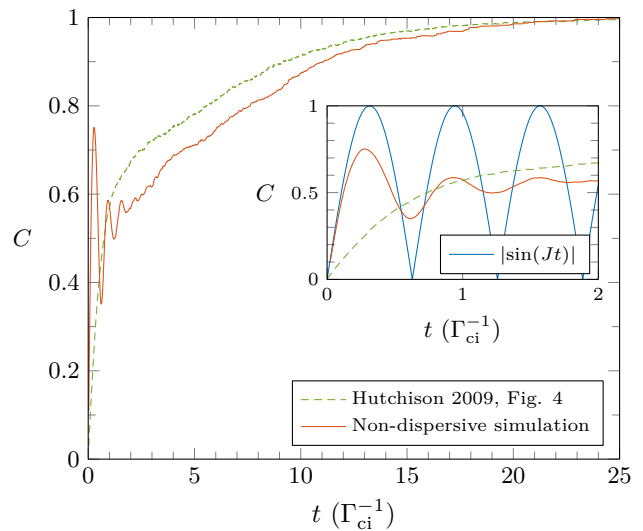

We can assess the performance of the entanglement procedure by examining \cref{fig:hutchFidelityVsSignal}.
Firstly, note the two dense clusters of points, one with high \(E_f\)  near \(\Vint = 0\) and the other with low \(E_f\)  near \(\Vint = -0.2\).
The first cluster has  \(E_f\) $\approx$ 1, indicating that a high maximum fidelity is achievable with this system and parameters. Additionally, the high- and low-fidelity clusters do not significantly overlap in terms of \(\Vint\). These simulations show that homodyne measurements of the cavity transmission provide meaningful information regarding the fidelity of the final qubit state to the target.

Shown in \cref{fig:hutchConcVsT} is a plot of the average concurrence
$C$ of the final state resulting from trajectories that satisfy \(\abs{\Vint\qty(t)} \leq t\sqrt{\Gamma_{\mathrm{ci}}}\).
This condition restricts the average to the trajectories within the high-fidelity cluster in \cref{fig:hutchFidelityVsSignal}.
The results presented in Ref.\ \cite{Hutchison2009} were obtained in the strongly dispersive regime, but the specific value of \(g/\Delta\) was neither relevant nor specified.  Due to the computational cost of extending the simulation time out to the resulting \(t_\text{final} \sim \Delta/g\), we opted to use a value of \(g/\Delta = 0.1\).

Note that our simulated values of concurrence in \cref{fig:hutchConcVsT} exhibit initial oscillatory behavior that is not observed in Ref.\ \cite{Hutchison2009}.
We attribute these oscillations to the cavity mediated interaction between the two qubits.
First, we note that
\begin{align}
	\alpha(t)=\frac{2 \ii p \sqrt{\kappa_{\mathrm{in}}}}{\kappa}+c_{\alpha} e^{-t\kappa/2}
	\label{eq:alphat}
\end{align}
is a solution of \cref{eqn:twoLevelAlphaDot}, for constant \(c_{\alpha}\) determined by the initial conditions.
For the value of \(\kappa\) taken in the \cref{fig:hutchFidelityVsSignal,fig:hutchConcVsT}, the second term of \cref{eq:alphat} decays quickly compared to the observed oscillation period.
Thus, we assume a time-independent displacement \(\alpha\). After transforming \cref{eqn:Hdisp} to the rotating reference frame defined by \cref{eqn:Utrans} with \(\omega_d=\omega_c\), we apply the displacement operator \(D\qty[\alpha]\) with constant \(\alpha\).
The dispersive Hamiltonian then takes the following form:

\begin{equation}
	\begin{split}
		H_{\text{disp}} ={}& \hbar
                \qty(\sum_{\mathclap{\hspace{2em}j\in\qty{A,B}}}\chi_j
                \varLabel{\sigmaZinit}{j})\qty(\actHat\aHat +
                \abs{\alpha}^2 - \conj\alpha\aHat - \alpha\actHat +\frac{1}{2})\\
		& + \hbar J\qty(\varLabel{\sigmaMinit}{A}\varLabel{\sigmaPinit}{B} + \varLabel{\sigmaPinit}{A}\varLabel{\sigmaMinit}{B}).
	\end{split}
	\label{eq:Hdisprotdisp}
      \end{equation}

Our choice of the displaced reference frame as defined in \cref{eqn:twoLevelAlphaDot} ensures that the cavity state is well approximated by the vacuum state, implying that the terms \(\propto \aHat,\actHat,\actHat\aHat\) in \cref{eq:Hdisprotdisp} can be neglected.
Employing this approximation and additionally requiring \(\chi_A=\chi_B=\chi\), the time evolution generated by \(H_{\text{disp}}\) is given by
\begin{align}
	U(t) ={}& \exp[-\frac\ii\hbar H_{\text{disp}}t]\\
	={}&
	\begin{aligned}[t]
		&\exp[-\ii\chi\qty(\frac{1}{2} + \abs{\alpha}^2)\varLabel{\sigmaZinit}{A}t]\\
		&\times\exp[-\ii\chi\qty(\frac{1}{2} + \abs{\alpha}^2)\varLabel{\sigmaZinit}{B}t]\\
		&\times\exp[-\ii J\qty(\varLabel{\sigmaMinit}{A}\varLabel{\sigmaPinit}{B}+\varLabel{\sigmaPinit}{A}\varLabel{\sigmaMinit}{B})t],
	\end{aligned}
	\label{eq:time_evolution}
\end{align}
where the commutation relations
\begin{align}
	\comm{\varLabel{\sigmaZinit}{A} + \varLabel{\sigmaZinit}{B}}{\varLabel{\sigmaMinit}{A}\varLabel{\sigmaPinit}{B} + \varLabel{\sigmaPinit}{A}\varLabel{\sigmaMinit}{B}} &= 0,\\
	\comm{\varLabel{\sigmaZinit}{A}}{\varLabel{\sigmaZinit}{B}} &= 0,
\end{align}
justify the breakdown of the exponential.
The first two exponentials in \cref{eq:time_evolution} describe single qubit rotations of qubits \(A\) and \(B\) about the \(z\)-axis, respectively, and do not affect the entanglement dynamics between subsystems \(A\) and \(B\).
The third exponential, however, acts on both qubits; its matrix representation with respect to the basis \(\qty{\ketSub{\downarrow}{A}\ketSub{\downarrow}{B},\ketSub{\downarrow}{A}\ketSub{\uparrow}{B},\ketSub{\uparrow}{A}\ketSub{\downarrow}{B},\ketSub{\uparrow}{A}\ketSub{\uparrow}{B}}\) is
\begin{align}
U_{2q}(t)&=\exp[-\ii J\qty(\varLabel{\sigmaMinit}{A}\varLabel{\sigmaPinit}{B}+\varLabel{\sigmaPinit}{A}\varLabel{\sigmaMinit}{B})t]\\
&=
\begin{bmatrix}
	1& 0 &0 &0 \\
	0& \phantom{-\ii}\cos(Jt) & -\ii\sin(Jt) &0\\
	0 & -\ii\sin(Jt)&\phantom{-\ii}\cos(Jt) &0 \\
	0&0&0&1
\end{bmatrix}.
\end{align}
This realizes a cavity-mediated entangling iSWAP gate for \(Jt = \qty(n + \frac12)\pi\) and \(n\in\mathbb{N}_0\), while leaving the system unchanged for \(Jt = n\pi\) \citep{Benito2019a,Young2022,Mielke2023}.
The initial state $\ket{\Psi_{\rm init}}$ evolves under the entangling interaction as
\begin{align}
	\ket{\psi_{2q}(t)} ={}& U_{2q}(t)\ketSub{+x}{A}\ketSub{+x}{B}\nonumber\\
	={}& \frac{1}{2}
	\begin{aligned}[t]
		\Bigl(&\ketSub{\downarrow}{A}\ketSub{\downarrow}{B}
		+ \qty(\cos(Jt) - \ii\sin(Jt))\ketSub{\downarrow}{A}\ketSub{\uparrow}{B}\\
		&+ \qty(\cos(Jt) - \ii\sin(Jt))\ketSub{\uparrow}{A}\ketSub{\downarrow}{B}
		+ \ketSub{\uparrow}{A}\ketSub{\uparrow}{B}\smash[t]{\Big)}.
  	\end{aligned}
\end{align}
The corresponding concurrence,
\begin{align}
	C\qty(\ketbra{\psi_{2q}(t)})=\abs{\sin(Jt)},
	\label{eqn:concOsc}
\end{align}
and the entanglement of formation,
\begin{multline}
	E_f\qty(\ketbra{\psi_{2q}(t)})\\
	=\frac{\abs{\cos(Jt)}\ln(\frac{2}{1+\abs{\cos(Jt)}} - 1)+\ln(\frac{4}{\sin[2](Jt)})}{\ln(4)},
\end{multline}
inherit the time periodicity of \(U_{2q}(t)\).

\Cref{fig:hutchConcVsT} demonstrates the accuracy of our simulations: the long-term behavior of the concurrence is a close match to that produced by the dispersive model of Hutchison \textit{et al.} \citep{Hutchison2009} and the oscillations observed at short times are consistent with \cref{eqn:concOsc}.

{
	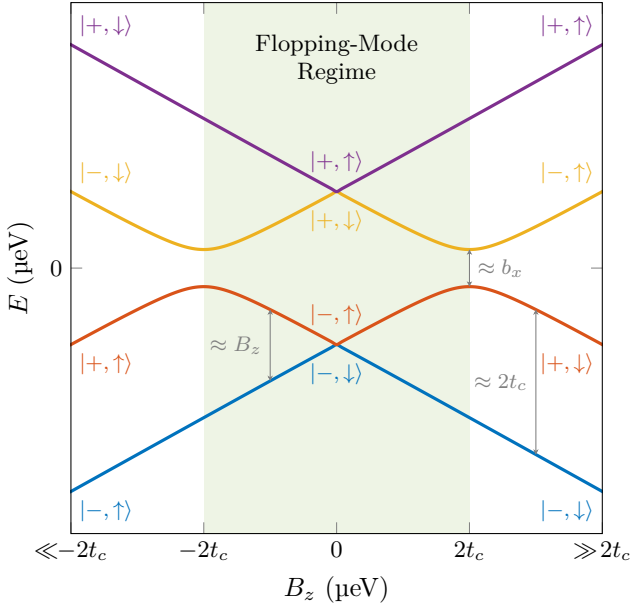
\begin{figure}[ht!]
		\centering
		\tikzsetnextfilename{LevelDiagram_new}
		\begin{tikzpicture}
			
		\end{tikzpicture}
		\caption{
			Energy level structure of a flopping-mode spin qubit with a micromagnet as a function of external magnetic field (\(\propto \wz\)) for fixed interdot tunnel coupling \(t_c\).
			The labels \(\ket{\mp,\downarrow\!\uparrow}\) show the corresponding charge-spin eigenstates in each regime.
			For the MBE protocols described in this paper the system is operated with \(\abs{\wz} < 2\tc\) (shaded in green).
	   		Here, the energy difference between the lowest two eigenstates is primarily spinlike, with a corresponding energy  \(\approx\wz\).
	   		The degree of spin-charge hybridization, and therefore the magnitude of the spin-photon coupling, increases as \(\abs{\wz} \rightarrow 2\tc\).
	   		In the regime where \(\wz\gg2\tc\), the transition between the two lowest eigenstates becomes primarily chargelike, with a corresponding transition energy \(2\tc\).
		}
		\label{fig:levelDiagram}
	\end{figure}
}

\section{\label{ch:DQD_MBE}Flopping-Mode Spin~Qubit Master~Equation}

We now consider two DQD flopping-mode spin qubits, again labeled \(A\) and \(B\), coupled to a superconducting coplanar waveguide resonator.
To derive the stochastic master equation for our MBE protocol, we begin with the (non-dispersive) two-qubit Tavis-Cummings Hamiltonian \citep{Tavis1968,Hutchison2009,Benito2017}, now with the full system Hamiltonian in the laboratory frame:
\begin{align}
	\label{eqn:HeffInit}
	\tilde{H} ={}& H_\text{cavity} + \sum_{\mathclap{j\in\qty{A,B}}}\pqty{\tilde{H}_\text{DQD}^\qty(j) + \tilde{H}_\text{inter}^\qty(j)},\\
	\label{eqn:HcavInit} H_\text{cavity} ={}& \hbar \wc\actHat\aHat + \hbar \sqrt{\kappa_\text{in}}\qty[\driveAmp\qty(t)\actHat\ee^{-\ii\wdr t}+\conj\driveAmp\qty(t)\aHat\ee^{\ii\wdr t}],\\
	\label{eqn:HqubitInit}\tilde{H}_\text{DQD}^\qty(j) ={}&  \frac{\epsilon_j}2\varLabel{\tauZinit}{j} + \varLabel{\tc}{j}\varLabel{\tauXinit}j + \frac{\varLabel{\wz}j}{2}\varLabel{\sigmaZinit}j + \frac{\varLabel{\wx}j}{2}\varLabel{\tauZinit}j\varLabel{\sigmaXinit}j,\\
	\label{eqn:HinterInit}\tilde{H}_\text{inter}^\qty(j) ={}& \hbar \varLabel{\gc}{j}\qty(\actHat + \aHat)\varLabel{\tauZinit}{j},
\end{align}
where \(\epsilon\) is the double-dot charge detuning, \(\tc\) is the inter-dot tunnel coupling, \(\wx\) is the energy associated with the transverse magnetic field difference between the dots, and \(\wz\) is the Zeeman energy due to the external magnetic field. The angular coupling frequency describing the interaction of the charge with the cavity electric field is \(\gc\). \(\tilde\tau_k\) and \(\sigma_k\) act on the charge (left or right dot occupation) and spin (up or down) degrees of freedom in \cref{eqn:HqubitInit}.

After applying the sequence of transformations specified in \cref{ch:hamiltDeriv}, we obtain the following effective Hamiltonian in the displaced and rotating frame:

\begin{equation}
	\begin{split}
		\label{eqn:HeffSymb}
		\hat{H} ={}& \hbar \Dc\actHat\aHat +  \sum_{j\in\qty{A,B}}\biggl[\frac{\hbar}{2}\varLabel{\Dch}j\varLabel{\tauZqubit}j + \frac{\hbar}{2}\varLabel{\Dsp}j\varLabel{\sigmaZqubit}j\\
		&+ \hbar \varLabel{g_c}j\sin(\varLabel{\theta_1}j + \varLabel{\theta_2}j)\varLabel{\tauZqubit}j\\
		&\quad\times\pqty{\pqty{\aHat - \alpha}\varLabel{\sigmaPqubit}j + \pqty{\actHat - \conj\alpha}\varLabel{\sigmaMqubit}j}\\
		&+\hbar \varLabel{g_c}j\cos(\varLabel{\theta_1}j + \varLabel{\theta_2}j)\\
		&\quad\times\pqty{\pqty{\aHat - \alpha}\varLabel{\tauPqubit}j + \pqty{\actHat - \conj\alpha}\varLabel{\tauMqubit}j}\biggr]
	\end{split}
\end{equation}
with the following definitions for each qubit,
\begin{gather}
	\theta_{\substack{\text{\tiny1}\\\text{\tiny2}}} = \tan[-1](\sqrt{\qty(\frac{2\tc\pm\wz}{\wx})^2 + 1} - \qty(\frac{2\tc\pm\wz}{\wx})),\\
	\omega_{\substack{\text{\tiny1}\\\text{\tiny2}}} = \frac{\wx}{2\hbar\sin(2\theta_{\substack{\text{\tiny1}\\\text{\tiny2}}})},\\
	\begin{aligned}
		\omega_{\substack{\text{\tiny ch}\\\text{\tiny sp}}} ={}& \omega_1 \pm \omega_2, &
		\Delta_{\substack{\text{\tiny ch}\\\text{\tiny sp}}} ={}& \omega_{\substack{\text{\tiny ch}\\\text{\tiny sp}}} - \wdr,
	\end{aligned}
\end{gather}
where \(\alpha\) is the cavity coherent state population determined by \cref{eqn:twoLevelAlphaDot}.
The Pauli operators \(\qty{\hat\tau,\,\hat\sigma}\) appearing in the DQD Hamiltonian
[\cref{eqn:HeffSymb}] act on the eigenbasis
\(\qty{\ket0, \ket1, \ket2, \ket3}\) with
\begin{equation}
	\label{eqn:eigDef}
	\begin{alignedat}{2}
		\ket{0}=& \cos\theta_1\ket{-,\downarrow} &&+ \sin\theta_1\ket{+,\uparrow},\\
		\ket{1}=& \cos\theta_2\ket{-,\uparrow} &&+ \sin\theta_2\ket{+,\downarrow},\\
		\ket{2}=& \cos\theta_2\ket{+,\downarrow} &&- \sin\theta_2\ket{-,\uparrow},\\
		\ket{3} =& \cos\theta_1\ket{+,\uparrow} &&- \sin\theta_1\ket{-,\downarrow},
	\end{alignedat}
\end{equation}
and the operators defined as
\begin{equation}
	\begin{alignedat}{9}
		\tauZqubit&&\{\ket0,&&\ket1,&&\ket2,&&\ket3\} &= \{&-\ket0,&&-\ket1,&&\ket2,&&\ket3&\},\\
		\tauMqubit&&\{\ket0,&&\ket1,&&\ket2,&&\ket3\} &= \{&0,&&0,&&\ket0,&&\ket1&\},\\
		\sigmaZqubit&&\{\ket0,&&\ket1,&&\ket2,&&\ket3\} &= \{&-\ket0,&&\ket1,&&-\ket2,&&\ket3&\},\\
		\sigmaMqubit&&\{\ket0,&&\ket1,&&\ket2,&&\ket3\} &= \{&0,&&\ket0,&&0,&&\ket2&\}.
	\end{alignedat}
	\label{eqn:newOpDefs}
\end{equation}

In the low field regime, i.e. $|B_z|<2t_c$ (shaded green area in Fig.~4), the ground state to first excited state transition energy is dominated by the Zeeman energy $B_z$, and the ground state to second excited state transition energy is $\approx 2t_c$.
 Consequently, we can assign `chargelike' and `spinlike' properties to the DQD eigenstates, i.e. \(\qty{\ket{\widetilde{-,\downarrow}},\ket{\widetilde{-,\uparrow}},\ket{\widetilde{+,\downarrow}},\ket{\widetilde{+,\uparrow}}}\) \citep{Benito2019a}.
The five terms in \cref{eqn:HeffSymb} then correspond approximately to the cavity, charge, and spin energies, and to spin-photon and charge-photon coupling, respectively.

Now that we have described the coherent evolution of the two-qubit-cavity system, we can construct the master equation describing the time evolution of the full system density operator \(\rho\) in the presence of charge and spin dephasing,
\begin{equation}
	\label{eqn:masterEq}
	\begin{split}
		\dd{\rho} &= \mathcal{L}_4\rho\dt\\
		&= -\frac{\ii}{\hbar}\comm{\hat H}{\rho}\dt\\
		&+\sum_{\mathclap{j\in\Bqty{A,B}\;}}\pqty{\frac{\varLabel{\gamma_\mathrm{sp}}{j}}{2}\lindOp{T\qty(\varLabel{\sigmaZinit}{j})}{\rho}+\frac{\varLabel{\gamma_\mathrm{ch}}{j}}{2}\lindOp{T\qty(\varLabel{\tauZcharge}{j})}{\rho}}\dt\\
		&+\kappa\qty\big(\lindOp{\aHat}{\rho})\dt + \frac{\kappa}{2}\comm{\alpha\actHat - \conj\alpha\aHat}{\rho}\dt,
	\end{split}
\end{equation}
where \(\varLabel{\gamma_{\mathrm{sp}}}{j}\), \(\varLabel{\gamma_{\mathrm{ch}}}{j}\), and \(\kappa\) are the spin dephasing, charge dephasing, and cavity decay rates, respectively, and \(T\qty(X)\) transforms an arbitrary operator \(X\) from the bonding-antibonding basis \(\qty{\ket{-\downarrow},\ket{-\uparrow},\ket{+\downarrow},\ket{+\uparrow}}\) to the DQD eigenbasis \(\qty{\ket0, \ket1, \ket2, \ket3}\).
The last term in \cref{eqn:masterEq} is a consequence of applying a displacement transformation to the cavity photon Lindblad term \(\kappa\lindOp{\aHat}{\rho}\) preceding it, and exactly cancels with the final term in \cref{eqn:HeffSymb}.
This cancellation is in fact the reason we added the term \(-\frac\kappa2\alpha\) to our choice of \(\dot\alpha\) in \cref{eqn:twoLevelAlphaDot}. Finally, we can obtain a stochastic master equation for homodyne readout using \cref{eqn:masterEqStoch}, replacing \(\mathcal L\) with \(\mathcal L_4\).

To calculate the fidelity and entanglement of formation of the final system state, we first trace out the cavity degree of freedom.
Since the resulting system consists of two four-level systems, we must also trace out the chargelike degrees of freedom for both DQDs to produce \(\rho_\text{red.} = \Tr_\text{ch}[\Tr_\text{cav.}[\rho_\text{cond.}]]\). While tracing out the cavity degree of freedom is straightforward, the trace over the chargelike degree of freedom requires some clarification: Denoting the spinlike qubit states by \(\ket{\widetilde{\zeta}}_j\) with \(j\in\{A,B\}\) and \(\zeta\in\{\downarrow,\uparrow\}\), the matrix elements of \(\rho_\text{red.}\) are defined by
\begin{multline}
		\braSub{\widetilde{\zeta}}{A}\braSub{\widetilde{\xi}}{B}\rho_{\text{red.}}\ketSub{\widetilde{\zeta^\prime}}{A}\ketSub{\widetilde{\xi^\prime}}{B} \\
	= \sum_{\mathclap{k,l\in\qty{-,+}}}{\braSub{\widetilde{k,\zeta}}{A}\braSub{\widetilde{l,\xi}}{B}\Tr_{\text{cav.}}\qty[\rho_{\text{cond.}}]\ketSub{\widetilde{k,\zeta^\prime}}{A}\ketSub{\widetilde{l,\xi^\prime}}{B}},
\end{multline}
for \(\zeta,\xi,\zeta^\prime,\xi^\prime \in \qty{\downarrow,\uparrow}\). The spin-charge hybridized states \(\ketSub{\widetilde{k,\zeta}}{j}\) can be identified with the DQD Hamiltonian eigenstates \(\{\ket{0},\ket{1},\ket{2},\ket{3}\}\) as explained above. Since \(\rho_\text{red.}\) is a \(4\times4\) density matrix corresponding to the spinlike states of the two DQDs, we can calculate \(F_{\ket{\Psi_+}}\) and \(F_e\) as we did for the two-level qubit model.

\section{\label{ch:SimResults}Results}

\begin{figure*}
	\tikzsetnextfilename{QuadPlotReplacedB}
	\begin{tikzpicture}
		
	\end{tikzpicture}
	\protect\caption
	{
		\label{fig:quadPlot}(a) Plot of \(F_{\ket{\Psi_+}}\) and $E_f$ as a function of time for groups of trajectories with the lowest, middle, and highest third of corresponding \(\Vint\) values, for the parameters listed in \cref{eqn:benitoParams4level}.
		(b) Scatter plot of \(F_{\ket{\Psi_+}}\) at \(t=\SI{100}{\us}\) as a function of $V_{\rm int}$ using the parameters given in \cref{eqn:benitoParams4level}.
		The vertical dashed lines correspond to the subgrouping of the data in (a).
		(c) Plot of \(F_{\ket{\Psi_+}}\) and $E_f$ as a function of time for the groups of trajectories with the lowest, middle, and highest third of corresponding \(\Vint\) values, for a cavity decay rate \(\kappa/2\pi\) = 15 MHz.
		All other parameters are the same as those given by \cref{eqn:benitoParams4level}.
		(d) Scatter plot of \(F_{\ket{\Psi_+}}\) at \(t=\SI{39}{\us}\) as a function of $V_{\rm int}$, for a cavity decay rate \(\kappa/2\pi\) = 15 MHz.
		All other parameters are the same as those given by \cref{eqn:benitoParams4level}.
	}
\end{figure*}
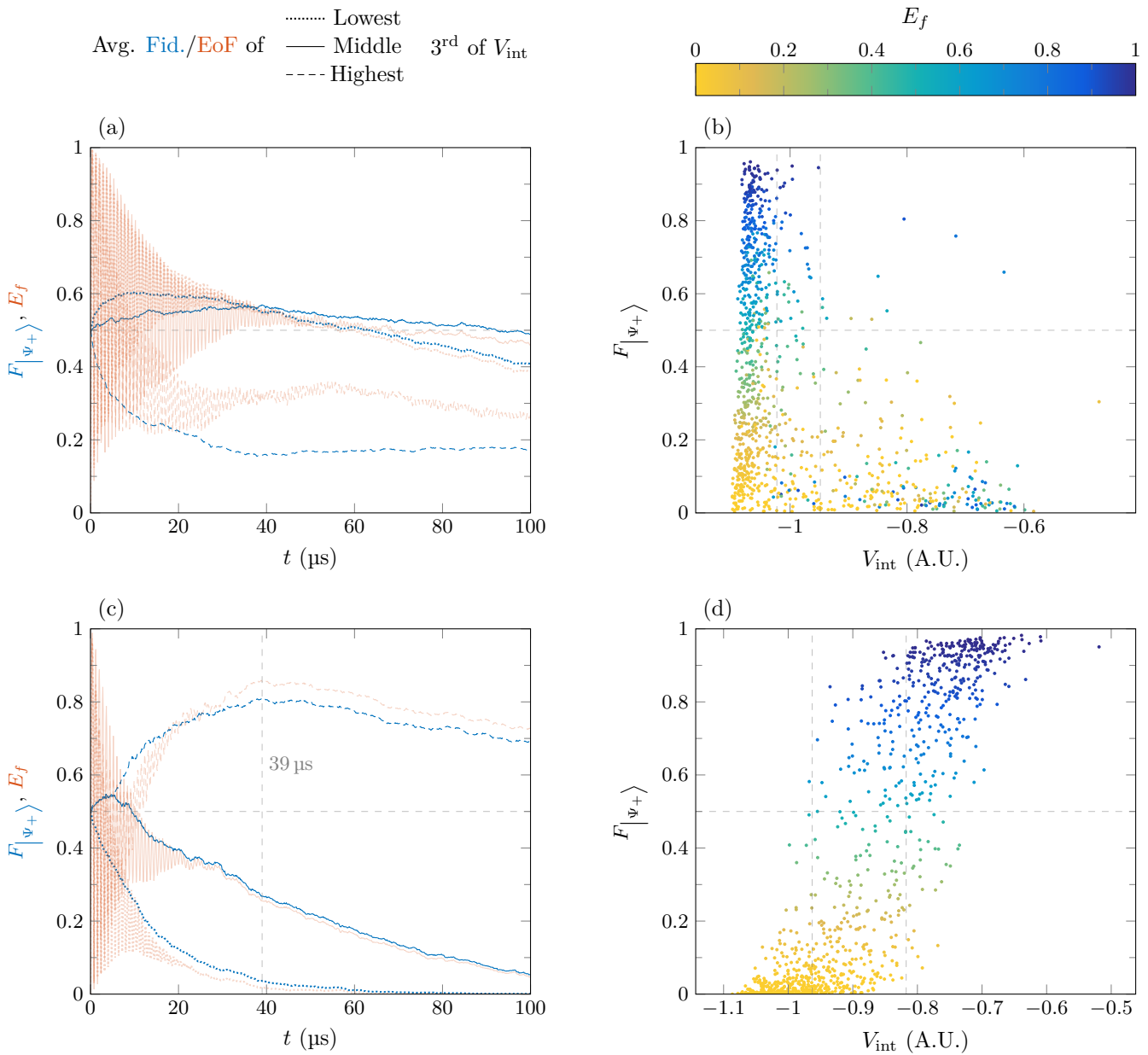

We can now simulate entanglement generation with parameters corresponding to spin-qubit cQED devices using the four-level model described by \cref{eqn:HeffSymb,eqn:masterEqStoch}. The parameters in \cref{eqn:benitoParams4level} are taken from a device designed to perform dispersive \emph{coherent} two-qubit operations \citep{Benito2019a}, i.e.\ a cavity iSWAP. The result of simulating 1000 unravelings of the stochastic master equation is shown in Figs.\ \ref{fig:quadPlot}(a--b). We can visualize the protocol's effect over time by taking the set of fidelities-to-target at each time and dividing them into three groups: one containing the third of trajectories with the lowest \(\Vint\), one containing the third with the highest \(\Vint\), and one containing the remaining trajectories. We can then plot the average fidelity-to-target for each of these groups over the duration of the measurement.
Such a plot is shown in \cref{fig:quadPlot}(a). Although the measurement achieves a fidelity to \(F_{\ket{\Psi_+}}\) greater than 50\% for the lowest and middle thirds of \(\Vint\) values, the inability to postselect for higher-fidelity states causes the achievable average fidelity to reach a maximum of \(\SI{\avgLowKappaFid}{\percent}\) for these device parameters.

\begin{table}[ht!]
	\centering
	\begin{tabular}{cl}
		\toprule
		Quantity & Definition \\
		\midrule
   		\(\ket{\psi_\text{initial}^\text{qubit}}\) & \(\frac{1}{2}\qty(\ket{0} + \ket{1})^{\otimes 2}\) \\
   		\(\ket{\psi_\text{initial}^\text{cavity}}\) & \(\ket{0}\) \\
   		\(g_c^\qty(A)/2\pi\) & \(\SI{50}{\MHz}\) \\
   		\(g_c^\qty(B)\) & \(-g_c^\qty(A)\) \\
   		\(\kappa/2\pi\) & \(\SI{1.5}{\MHz}\) \\
   		\(t_c\) & \(\SI{13.2}{\micro\electronvolt}\) \\
   		\(B_z\) & \(\SI{24}{\micro\electronvolt}\) \\
   		\(b_x\) & \(\SI{2}{\micro\electronvolt}\) \\
   		\(\phi\) & \(\SI{0}{\radian}\) \\
   		\(\eta\) & \(1\) \\
   		\(\alpha\qty(\forall t)\) & \(0.1\ii\) \\
   		\(\gamma_{\mathrm{ch}}/2\pi\) & \(\SI{2.5}{\MHz}\) \\
   		\(\gamma_{\mathrm{sp}}/2\pi\) & \(\SI{0.005}{\MHz}\)\\
   		\(N_\text{trajs.}\) & \(\num{1000}\) \\
   		\(\wdr\) & \(\wc\) \\
   		\(\Dsp\) & \(10g_c^\qty(A)\)\\
   		\(\kappa_{\text{in}}/\kappa_{\text{out}}\) & \(1/8\)\\
   		\bottomrule
	\end{tabular}
	\caption
	{
		\label{eqn:benitoParams4level}
		DQD device parameters taken from \citealp[Ref.][]{Benito2019a}.
	}
\end{table}

\Cref{fig:quadPlot}(b) shows a scatter plot \(F_{\ket{\Psi_+}}\) as a function of $V_{\rm int}$ with a measurement time of \(\SI{100}{\us}\). The trajectories have a wide range of fidelities between \(\SI{0}{\percent}\) and \(\SI{95}{\percent}\). A correlation between \(\Vint\) and \(F_{\ket{\Psi_+}}\) is evident; however, the slope of this correlation is much too high to reliably postselect for high-fidelity states, given the nonzero spread in \(\Vint\) for trajectories of equal fidelity. In principle we could simply increase the measurement time further, allowing the trajectories to form high- and low-fidelity clusters with a resolvable \(\Vint\) separation. However, the spin dephasing time \(2\pi/\gamma_{\mathrm{sp}}=\SI{200}{\us}\) establishes an upper bound for usable measurement times, since states measured over a time \(t\sim\SI{200}{\us}\) will be subject to significant spin dephasing before the measurement is complete. 

One approach to improve MBE performance is to increase the cavity outcoupling rate \(\kappa\), which will increase the magnitude of the stochastic term in \cref{eqn:masterEqStoch}, and therefore increase the relative effect of the measurement on the final qubit state. In Figs.\ \ref{fig:quadPlot}(c--d), we examine the result of a \(10\times\) increase in \(\kappa\). The increase in \(\kappa\) results in a significant enhancement in the overall MBE protocol fidelity: the mean fidelity of the trajectories in the highest third of integrated homodyne outputs \(\Vint\) is  \(F_{\ket{\Psi_+}}\) = \(\SI{\avgHighKappaFid}{\percent}\), as shown in \cref{fig:quadPlot}(c). \Cref{fig:quadPlot}(d) provides additional information on the relationship between \(F_{\ket{\Psi_+}}\), \(\Vint\), and  $F_e$ at the final measurement time of \(\SI{39}{\us}\).
In the upper right portion of \cref{fig:quadPlot}(d), we observe the cluster of trajectories that have been projected onto the target \(\ket{\Psi_+}\), with a peak $E_f$ = \(\SI{\maxHighKappaFid}{\percent}\) and an average $E_f$ = \(\SI{\avgHighKappaFid}{\percent}\) if postselecting the highest one-third of homodyne measurement results.

\begin{figure}[t!]
	\centering
	\tikzsetnextfilename{ThreshPlot_new}
	\begin{tikzpicture}
		
	\end{tikzpicture}
	\caption{Plot of the average \(F_{\ket{\Psi_+}}\) (blue) and \(E_f\) (orange), when postselecting the \(Y\) \% of trajectories with the highest values of \(\Vint\). These simulations correspond to \(\kappa/2\pi\) = 15 MHz and a measurement time of \(\SI{39}{\us}\), i.e. the data in \cref{fig:quadPlot}(d).}
	\label{fig:threshPlot}
\end{figure}
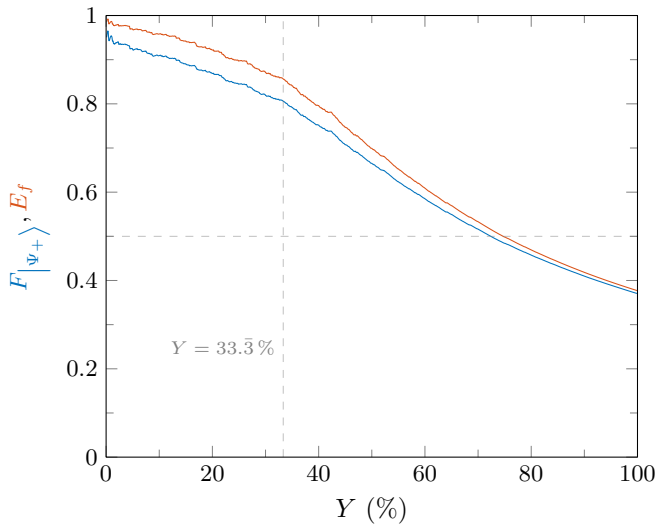

\Cref{fig:threshPlot} shows how the average \(F_{\ket{\Psi_+}}\) and the average $E_f$ respond to changes in the postselection yield. For the data in 
\cref{fig:quadPlot}(d) we select trajectories with \(\Vint\) \emph{greater than} a threshold. We
use the postselection yield \(Y\), the fraction of trajectories satisfying the postselection criteria, in place of the associated maximum \(\Vint\).
The solid lines in \cref{fig:threshPlot} present equivalent information corresponding to a device with an increased outcoupling rate \(\kappa/2\pi\) = 15 MHz, at time \(t = \SI{39}{\us}\).
As the postselection yield is reduced, the average fidelity of the postselected states rises, eventually reaching a maximum value of \(\SI{\sim95}{\percent}\).
The entanglement of formation similarly increases to a maximum value of \(\SI{\sim99}{\percent}\) for low yields.
A more realistic postselection yield of 33\% can, as mentioned previously, achieve reasonable fidelities of \(\SI{\sim\avgHighKappaFid}{\percent}\); this yield is indicated by the dashed vertical line in \cref{fig:threshPlot}.

 Overall, if we are willing to discard a sufficient fraction of measurement runs, increasing the cavity outcoupling rate to $\kappa/2\pi$ = 15 MHz results in significant improvements to the average Bell-state fidelity and entanglement-of-formation of the postselected two-qubit states. Based on these observations, we predict that a DQD-based cQED device with \(\kappa/2\pi\) = 15 MHz, and other parameters specified by \cref{eqn:benitoParams4level}, would be a capable of a demonstration of MBE.

\section{\label{ch:Conclusion}Conclusions}
We have developed a theoretical model to simulate the evolution of two silicon flopping mode spin qubits coupled to a microwave resonator and subjected to a continuous homodyne-based parity measurement.
Results from these simulations suggest that current Si/SiGe DQD cavity devices, designed to utilize photon-mediated spin-spin coupling, would only be able to achieve an entanglement fidelity of \(\SI{\sim\avgLowKappaFid}{\percent}\) when subjected to realistic levels of charge and spin dephasing.
A device better suited to such an experiment would likely need to incorporate an output cavity coupling rate \(\sim\)10\(\times\) higher than current devices. Simulations indicate that with such a cavity, the device would be able to achieve entanglement fidelities of \(\SI{\avgHighKappaFid}{\percent}\) at a predicted 33~\% postselection success probability, based on a homodyne measurement of the cavity output.

\begin{acknowledgments}
Supported by Army Research Office grants W911NF-15-1-0149 and W911NF-23-1-0104.
\end{acknowledgments}

\appendix

\section{\label{ch:hamiltDeriv}Detailed Derivation of the Effective Hamiltonian}

As the qubits have no direct interactions with each other, for the moment we will omit the sum over the two qubits and examine the Hamiltonian of a single qubit coupled to the cavity. Since we do not make a dispersive approximation, we will not perform any changes of basis that hybridize qubit and cavity states; obtaining the two-qubit Hamiltonian from the single-qubit version is therefore straightforward.

In the case of zero charge detuning (\(\epsilon=0\)), the eigenbasis of the charge-qubit Hamiltonian \(H_\text{charge} = \tc\tauXinit\) is simply the bonding-antibonding basis, \(\ket{\mp} = \frac{1}{\sqrt{2}}\qty(\ket{R} \mp \ket{L})\).

Writing our charge-state Pauli operators as \(\tau\) after transforming to this basis, we find that \(\tauZinit\rightarrow\tauXcharge\) and \(\tauXinit\rightarrow\tauZcharge\).
Therefore, our Hamiltonian from \cref{eqn:HeffInit} becomes:
\begin{align}
	\label{HeffCharge}H ={}& H_\text{cavity} + H_\text{DQD} + H_\text{inter},\\
	\label{eqn:HqubitCharge} H_\text{DQD} ={}& t_c\tauZcharge + \frac{\wz}{2}\sigmaZinit + \frac{\wx}{2}\tauXcharge\sigmaXinit,\\
  \label{eqn:HinterCharge} H_\text{inter} ={}& \hbar g_c\qty(\actHat + \aHat)\tauXcharge,
\end{align}
with \(H_\text{cavity}\) as defined in \cref{eqn:HcavInit}.

As shown in \cref{eqn:HqubitCharge}, the electron spin states couple to the electron charge states via a micromagnet-induced magnetic field gradient, which produces the spin-dependent interdot energy difference \(\wx\).
The DQD charge states in turn couple to the cavity field via a typical dipole field term, shown in \cref{eqn:HinterCharge}.

We now perform another change of basis, this time to diagonalize \(H_\text{DQD}\) as given in \cref{eqn:HqubitCharge}.
After diagonalizing, we are able to define \(\omega_1\), \(\omega_2\) such that the eigenfrequencies of \(H_\text{DQD}\) are \(\qty{-\omega_1,\,-\omega_2,\,\omega_2,\,\omega_1}\).

In the DQD eigenbasis given in \cref{eqn:eigDef}, our Hamiltonian becomes
\begin{align}
	\hat{H} ={}& H_\text{cavity} + \hat{H}_\text{qubit} + \hat{H}_\text{inter},\\
	\hat{H}_\text{DQD} ={}& \hbar \omega_1\qty(\hat\Pi_3 -
                                \hat\Pi_0) + \hbar \omega_2\qty(\hat\Pi_2 - \hat\Pi_1),\\
  \label{eqn:HinterQubit} \hat{H}_\text{inter} ={}& \hbar
                                                   \gc\qty(\actHat + \aHat)\tauXcharge,
\end{align}
where $\hat\Pi_k=|k\rangle\langle k|$ is the projector onto DQD eigenstate \(\ket{k}\) (\(k\in\{0,1,2,3\}\)) given in \cref{eqn:eigDef}, and \(H_\text{cavity}\) is as defined in \cref{eqn:HcavInit}. Transforming \(\tauXcharge\) to the DQD eigenbasis, we find
\begin{align}
	\tau_x\rightarrow \cos(\theta_1+\theta_2)\hat{\tau}_x+\sin(\theta_1+\theta_2)\hat{\tau}_z\hat{\sigma}_x,
\end{align}
where the Pauli operators \(\hat{\tau}_i\)  and \(\hat{\sigma}_i\)  are defined in \cref{eqn:newOpDefs}.

If we additionally transform to a rotating frame via the transformation \(V=\ee^{-\ii \hat{H}_\text{DQD}t/\hbar}\), the DQD Hamiltonian vanishes, while inside the interaction Hamiltonian, \(\tauXcharge\) becomes
\begin{multline}
\label{eqn:tauXqubitRot}
	\cos(\theta_1+\theta_2)\qty(\ee^{-\ii\qty(\omega_1+\omega_2)t}\tauMqubit+ \ee^{\ii\qty(\omega_1+\omega_2)t}\tauPqubit)\\
	+\sin(\theta_1+\theta_2)\tauZqubit\qty(\ee^{-\ii\qty(\omega_1-\omega_2)t}\sigmaMqubit+ \ee^{\ii\qty(\omega_1-\omega_2)t}\sigmaPqubit).
\end{multline}
Due to the association of the quantities \(\omega_1+\omega_2\) and \(\omega_1-\omega_2\) with charge and spin transitions, respectively, we define \(\wch=\omega_1+\omega_2\) and \(\wsp=\omega_1-\omega_2\).
These are the `chargelike' and `spinlike' frequencies of the hybridized qubit.

We now turn our attention to the cavity Hamiltonian given in \cref{eqn:HcavInit}.
To eliminate the terms in this equation proportional to \(\driveAmp\qty(t)\), we first transform to a rotating reference frame.
The transformation is specified by the operator \(W=\ee^{-\ii\wdr\actHat\aHat t}\).
Applying $W$ to \cref{eqn:HcavInit}, we obtain,
\begin{equation}
	H_\text{cavity}^\prime = \smash[t]{\hbar\overbrace{\qty(\wc - \wdr)}^{\equiv\Delta_c }}\actHat\aHat + \hbar \sqrt{\kappa_{\rm in}}(\driveAmp\qty(t)\actHat + \conj{\driveAmp}\qty(t)\aHat ).
\end{equation}

We then apply a displacement transformation \(D\qty[\alpha]\coloneqq\ee^{\alpha\actHat - \conj\alpha\aHat}\), obtaining:
\begin{equation}
	\begin{split}
		\hat{H}_\text{cavity} ={}& \hbar
                \Delta_c\qty(\actHat\aHat + \abs{\alpha}^2) \\
                & - \hbar \sqrt{\kappa_{\rm in}} \qty(\driveAmp\conj\alpha + \conj\driveAmp\alpha)
                + \frac{\ii \hbar}{2}\qty(\alpha\conj{\dot\alpha} - \conj\alpha\dot\alpha)\\
		{}& + \hbar \qty(\sqrt{\kappa_{\rm in}} \driveAmp - \Delta_c\alpha + \ii\dot\alpha)\actHat\\
		{}& +\hbar (\sqrt{\kappa_{\rm in}} \conj\driveAmp - \Delta_c\conj\alpha - \ii\conj{\dot\alpha})\aHat.
	\end{split}
	\label{eqn:HcavDisp}
\end{equation}
An intuitive choice for \(\alpha\) would be \(\dot\alpha \equiv -\ii\Delta_c\alpha + \ii \sqrt{\kappa_{\rm in}} \driveAmp\qty(t)\), which would nullify the final two terms in \cref{eqn:HcavDisp}.
This would be appropriate if simplifying the Hamiltonian were our only concern.
With the benefit of hindsight, we instead select \(\dot\alpha \equiv -\ii\Delta_c\alpha + \ii \sqrt{\kappa_{\rm in}} \driveAmp\qty(t) -\frac{\kappa}{2}\alpha\), for some as yet undefined \(\kappa\).
With this choice, \cref{eqn:HcavDisp} reduces to
\begin{equation}
	\hat{H}_\text{cavity} = \hbar \Delta_c\actHat\aHat -\frac{\hbar}{2}\sqrt{\kappa_{\rm in}}\qty(\driveAmp\conj\alpha + \conj\driveAmp\alpha) - \frac{\ii\hbar}{2}\kappa\qty(\alpha\actHat - \conj\alpha\aHat).
	\label{eqn:HcavDisp2}
\end{equation}
The second term in \cref{eqn:HcavDisp2} is a state-independent constant and can therefore be discarded.

We now apply the cavity rotation and displacement transformations to the interaction term in \cref{eqn:HinterQubit}.
From this we obtain
\begin{equation}
	\label{eqn:HeffPreRW}
	\begin{split}
		\hat{H} ={}& \hbar\Delta_c\actHat\aHat + \hbar g_c\qty(\ee^{\ii\omega_d t}\qty(\actHat - \conj\alpha) + \ee^{-\ii\omega_d}\qty(\aHat - \alpha))\tauXcharge\\
		{}&-\frac{\ii\hbar}{2}\kappa\qty(\alpha\actHat - \conj\alpha\aHat),
	\end{split}
\end{equation}
with \(\tauXcharge\) taking the form given in \cref{eqn:tauXqubitRot}.

Finally, we can make a rotating-wave approximation, assuming that \(\qty{\wch-\wdr,\wsp-\wdr} \ll \qty{\wch+\wdr,\wsp+\wdr}\).
This allows us to eliminate rapidly oscillating sub-terms from the second term of \cref{eqn:HeffPreRW}, to obtain:
\begin{equation}
	\label{eqn:Heff2}
	\begin{split}
		\hat H ={}& \hbar \Dc\actHat\aHat\\
 		&+\hbar\gc
 		\begin{aligned}[t]
 			\biggl\{&\pqty{\actHat-\conj{\alpha}}
 			\begin{aligned}[t]
 				\Bigl[&\ee^{-\ii\Dch t}\cos(\theta_1 + \theta_2)\tauMqubit\\
 				&+\ee^{-\ii\Dsp t}\sin(\theta_1 + \theta_2)\tauZqubit\sigmaMqubit\Bigr]
 			\end{aligned}\\
			&+\text{h.c.}\biggr\}-\frac{\ii\hbar}{2}\kappa\qty(\alpha\actHat - \conj\alpha\aHat),
		\end{aligned}
	\end{split}
\end{equation}
where \(\Dch = \wch - \wdr\) and \(\Dsp = \wsp - \wdr\).

If we apply two additional rotating frame transformations, defined by \(V_\text{ch} = \ee^{\frac{1}{2}\ii\Dch\tauZqubit t}\) and \(V_\text{sp} = \ee^{\frac{1}{2}\ii\Dsp\sigmaZqubit t}\), we obtain \cref{eqn:HeffSymb}, a Hamiltonian that contains no explicit time-dependence once its final term cancels with the corresponding term in \cref{eqn:masterEq}.

\bibliographystyle{apsrev4-2}
\bibliography{MBErefs_v4}

\end{document}